\RequirePackage{lineno}
\documentclass[aps,prl,twocolumn,showpacs,groupedaddress,showkeys,floatfix]{revtex4}
\usepackage{amsmath}    
\usepackage{amssymb}    
\usepackage{bm}    
\usepackage{color}      
\usepackage{subfigure}  
\usepackage{epsfig} 
\usepackage{graphicx} 
\usepackage{color}
\usepackage{amsmath, amsthm}
\begin{document}


\newcommand{\D}{\displaystyle} 
\newcommand{\T}{\textstyle} 
\newcommand{\SC}{\scriptstyle} 
\newcommand{\SSC}{\scriptscriptstyle} 
\newcommand{\aha}[1]{\textcolor{magenta}{#1}}

\def\AJ{{ Astron. J.} } \def\ARAA{{ Ann. Rev. Astron. \& Astrophys.} }
\def\ApJ{{ Astrophys. J.} } \def\ApJL{{ Astrophys. J. Lett.} }
\def\ApJS{{ Astrophys. J. Suppl.} } \def\ApP{{ Astropart. Phys.} }
\def\AA{{ Astron. \& Astroph.} } \def\AAR{{ Astron. \& Astroph. Rev.}
} \def\AAL{{ Astron. \& Astroph. Lett.} } \def\AASu{{ Astron. \&
    Astroph. Suppl.} } \def\AN{{ Astron. Nachr.} } \def\IJMP{{
    Int. J. of Mod. Phys.} } \def\JCAP{{\it Journ. of Cosmol. \&
    Astropart. Phys.}Ê} \def\JGR{{ Journ. of Geophys. Res.}}
\def\JHEP{{ Journ. of High En. Phys.} } \def\JPhG{{ Journ. of Physics}
  {\bf G} } \def\MNRAS{{ Month. Not. Roy. Astr. Soc.} } \def\Nature{{
    Nature} } \def\NewAR{{ New Astron. Rev.} } \def\PASP{{
    Publ. Astron. Soc. Pac.} } \def\PhFl{{ Phys. of Fluids} }
\def\PLB{{ Phys. Lett.}{\bf B} } \def\PR{{ Phys. Rev.} } \def\PRD{{
    Phys. Rev.} {\bf D} } \def\PRL{{ Phys. Rev. Lett.} } \def\RMP{{
    Rev. Mod. Phys.} } \def\Science{{ Science} } \def\ZfA{{
    Zeitschr. f{\"u}r Astrophys.} } \def\ZfN{{ Zeitschr. f{\"u}r
    Naturforsch.} } \def\etal{{ et al.}}

\hyphenation{mono-chro-matic sour-ces Wein-berg chang-es Strah-lung
  dis-tri-bu-tion com-po-si-tion elec-tro-mag-ne-tic ex-tra-galactic
  ap-prox-i-ma-tion nu-cle-o-syn-the-sis re-spec-tive-ly su-per-nova
  su-per-novae su-per-nova-shocks con-vec-tive down-wards es-ti-ma-ted
  frag-ments grav-i-ta-tion-al-ly el-e-ments me-di-um ob-ser-va-tions
  tur-bul-ence sec-ond-ary in-ter-action in-ter-stellar spall-ation
  ar-gu-ment de-pen-dence sig-nif-i-cant-ly in-flu-enc-ed par-ti-cle
  sim-plic-i-ty nu-cle-ar smash-es iso-topes in-ject-ed in-di-vid-u-al
  nor-mal-iza-tion lon-ger con-stant sta-tion-ary sta-tion-ar-i-ty
  spec-trum pro-por-tion-al cos-mic re-turn ob-ser-va-tion-al
  es-ti-mate switch-over grav-i-ta-tion-al super-galactic com-po-nent
  com-po-nents prob-a-bly cos-mo-log-ical-ly Kron-berg Berk-huij-sen Karls-ru-he}
\def\simle{\lower 2pt \hbox {$\buildrel < \over {\scriptstyle \sim
    }$}} \def\simge{\lower 2pt \hbox {$\buildrel > \over {\scriptstyle
      \sim }$}} \def\intunits{{\rm s}^{-1}\,{\rm sr}^{-1} {\rm
    cm}^{-2}}


\title{Kneelike structure in the spectrum of the heavy component of cosmic rays 
observed with KASCADE-Grande} 

\author{
W.D.~Apel$^{1}$,
J.C.~Arteaga-Vel\'azquez$^{2}$,
K.~Bekk$^{1}$,
M.~Bertaina$^{3}$,
J.~Bl\"umer$^{1,4}$,
H.~Bozdog$^{1}$,
I.M.~Brancus$^{5}$,
P.~Buchholz$^{6}$,
E.~Cantoni$^{3,7}$,
A.~Chiavassa$^{3}$,
F.~Cossavella$^{4,}$\footnote{now at: Max-Planck-Institut f\"ur Physik, M\"unchen, Germany},
K.~Daumiller$^{1}$,
V.~de Souza$^{8}$,
F.~Di~Pierro$^{3}$,
P.~Doll$^{1}$,
R.~Engel$^{1}$,
J.~Engler$^{1}$,
M. Finger$^{4}$, 
D.~Fuhrmann$^{9}$,
P.L.~Ghia$^{7}$, 
H.J.~Gils$^{1}$,
R.~Glasstetter$^{9}$,
C.~Grupen$^{6}$,
A.~Haungs$^{1}$,
D.~Heck$^{1}$,
J.R.~H\"orandel$^{10}$,
D.~Huber$^{4}$,
T.~Huege$^{1}$,
P.G.~Isar$^{1,}$\footnote{now at: Institute of Space Sciences, Bucharest, Romania},
K.-H.~Kampert$^{9}$,
D.~Kang$^{4}$, 
H.O.~Klages$^{1}$,
K.~Link$^{4}$, 
P.~{\L}uczak$^{11}$,
M.~Ludwig$^{4}$,
H.J.~Mathes$^{1}$,
H.J.~Mayer$^{1}$,
M.~Melissas$^{4}$,
J.~Milke$^{1}$,
B.~Mitrica$^{5}$,
C.~Morello$^{7}$,
G.~Navarra$^{3,}$\footnote{deceased},
J.~Oehlschl\"ager$^{1}$,
S.~Ostapchenko$^{1,}$\footnote{now at: University of Trondheim, Norway},
S.~Over$^{6}$,
N.~Palmieri$^{4}$,
M.~Petcu$^{5}$,
T.~Pierog$^{1}$,
H.~Rebel$^{1}$,
M.~Roth$^{1}$,
H.~Schieler$^{1}$,
F.G.~Schr\"oder$^{1}$,
O.~Sima$^{12}$,
G.~Toma$^{5}$,
G.C.~Trinchero$^{7}$,
H.~Ulrich$^{1}$,
A.~Weindl$^{1}$,
J.~Wochele$^{1}$,
M.~Wommer$^{1}$,
J.~Zabierowski$^{11}$\\ 
\noaffiliation{(KASCADE-Grande Collaboration)}
}
\affiliation{
$^1$ Institut f\"ur Kernphysik, KIT - Karlsruher Institut f\"ur Technologie, Germany\\
$^2$ Universidad Michoacana, Instituto de F\'{\i}sica y Matem\'aticas, Morelia, Mexico\\
$^3$ Dipartimento di Fisica Generale dell' Universit\`a Torino, Italy\\
$^4$ Institut f\"ur Experimentelle Kernphysik, KIT - Karlsruher Institut f\"ur Technologie, Germany\\
$^5$ National Institute of Physics and Nuclear Engineering, Bucharest, Romania\\
$^6$ Fachbereich Physik, Universit\"at Siegen, Germany\\
$^7$ Istituto di Fisica dello Spazio Interplanetario, INAF Torino, Italy\\
$^8$ Universidade S\~ao Paulo, Instituto de F\'{\i}sica de S\~ao Carlos, Brasil\\
$^9$ Fachbereich Physik, Universit\"at Wuppertal, Germany\\
$^{10}$ Department of Astrophysics, Radboud University Nijmegen, The Netherlands\\
$^{11}$ National Centre for Nuclear Research, Lodz, Poland\\
$^{12}$ Department of Physics, University of Bucharest, Bucharest, Romania\\
}
\date{\today}
\email{haungs@kit.edu}

\begin{abstract}
We report the observation of a steepening in the cosmic ray energy spectrum 
of heavy primary particles at about $8 \cdot 10^{16}\,$eV. 
This structure is also seen in the all-particle energy spectrum, but is less significant.
Whereas the `knee' of the cosmic ray spectrum at $3$-$5 \cdot 10^{15}\,$eV
was assigned to light primary masses by the KASCADE experiment, the new structure 
found by the KASCADE-Grande experiment is caused by heavy primaries. 
The result is obtained by independent measurements of the charged particle 
and muon components of the secondary particles of extensive air showers in 
the primary energy range of $10^{16}$ to $10^{18}\,$eV. 
The data are analyzed on a single-event basis taking into account also the 
correlation of the two observables. 
\end{abstract}


\maketitle

The determination of the primary energy and composition in the energy range 
from $10^{15}\,$eV up to above $10^{20}\,$eV has been subject of earthbound experiments 
for more than five decades. 
It has been shown that the high-energy all-particle spectrum has a power-law 
like behavior ($\propto E^{\gamma}$, $\gamma \approx -2.7$), 
with features known as the `knee' at $3$-$5 \cdot 10^{15}\,$eV 
and the `ankle' at $4$-$10 \cdot 10^{18}\,$eV, respectively. 
Whereas at the knee the spectrum steepens ($\Delta \gamma = -\,0.3$-$0.4$), the ankle is 
characterized by a flattening of the spectrum ($\Delta \gamma = +\,0.3$-$0.4$). 
The KASCADE experiment has shown that the knee is due to a distinct decrease 
in the flux of primaries with light mass ($Z<6$)~\cite{kas-muo,kas-unf}. 
Many astrophysical models commenting on the origin of the knee assume a 
dependence of such break-offs (`knees') on the charge of the 
primary nuclei~\cite{peters,hoera}. 
Assuming that the knee is related to a break in the spectrum of primary 
Hydrogen nuclei, a knee-like structure in the spectrum of the heavy component 
($Z>13$ up to Iron nuclei) is expected in the energy range from 
about $4 \cdot 10^{16}\,$eV to about $1.2 \cdot 10^{17}\,$eV. 
So far, such a structure has not been observed experimentally.
We present measurements of extensive air showers (EAS) in the primary 
energy range of $10^{16}\,$eV to $10^{18}\,$eV performed with KASCADE-Grande 
(KArlsruhe Shower Core and Array DEtector with Grande extension) and investigate 
the mass composition of the cosmic rays. 
 \begin{figure}[!t]
  \vspace{5mm}
  \centering
  \includegraphics[width=2.1in]{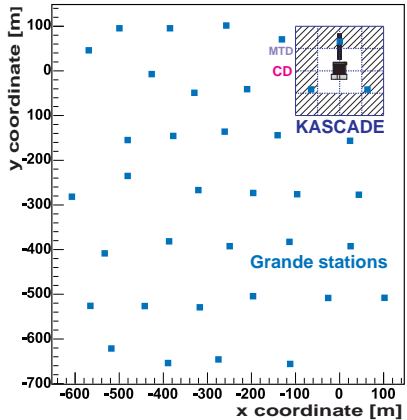}
	\caption{Layout of the KASCADE-Grande experiment: shown are the Grande array 
	as well as the KASCADE array with its central detector (CD) and 
	muon tracking detector (MTD). 
	The shaded area marks the outer 12 clusters (16 detector stations each) 
	of the KASCADE array consisting of shielded (muon array) and unshielded detectors. 
	The inner 4 clusters consist of unshielded detectors, only.}
	\label{fig1}
 \end{figure}
%
 
KASCADE-Grande, located at 49.1$^\circ$N, 8.4$^\circ$E, 110$\,$m$\,$a.s.l., consists of
the Grande array with 37 stations of 10$\,$m$^2$ scintillation detectors each, 
spread over an area of $700 \times 700\,$m$^2$, the original KASCADE
array covering $200 \times 200\,$m$^2$ with unshielded and shielded detectors, 
a muon tracking device, and a large calorimeter~\cite{kg-NIM10,kascade}. 
This multi-detector system allows us to investigate in detail the EAS 
generated by high-energy primary cosmic rays in the atmosphere.
For the present analysis, the estimation of energy and mass of the primary 
particles is based on the combined measurement of the charged particle component 
by the detector array of Grande and the muon component 
by the KASCADE muon array (Fig.~\ref{fig1}).
Basic shower observables like the core position, zenith angle, 
and total number of charged particles (shower size $N_{ch}$) are derived from
the measurements of the Grande stations. 
While the Grande detectors are sensitive to charged particles, the muonic 
component is measured independently by the shielded detectors of the KASCADE array.
192 scintillation detectors of $3.24\,$m$^2$ sensitive area each are placed below an iron 
and lead absorber to select muons above 230 MeV kinetic energy.  
A core position resolution of $5\,$m, a direction resolution of 
$0.7^\circ$, and a resolution of the shower size of about $15$\% are achieved.  
The total number of muons ($N_\mu$) with a resolution of about $25$\%
is calculated by combining the core position determined by the Grande 
array and the muon densities measured at the KASCADE array, where 
$N_\mu$ undergoes a correction for a bias in reconstruction due 
to the asymmetric position of the detectors~\cite{kg-NIM10}.
 \begin{figure}[!t]
  \vspace{5mm}
  \centering
  \includegraphics[width=2.9in]{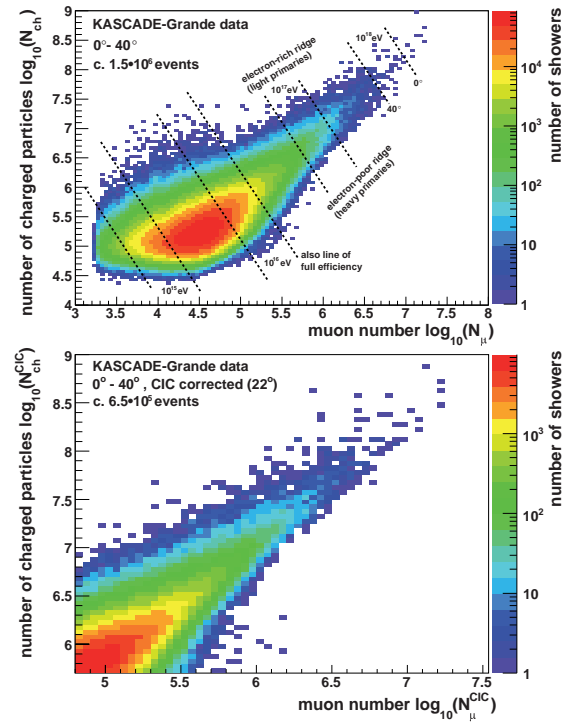}
	\caption{Two-dimensional distribution of the shower sizes: charged particle number 
	and total muon number. All quality cuts are applied. 
	In addition, a roughly estimated energy scale is indicated in the upper panel. 
	The lower panel shows a zoom to higher energies of the same observables, 
	now corrected for attenuation.}
	\label{2dim}
 \end{figure}
%

The present analysis is based on 1173 days of data taking. The cuts on the sensitive 
area (EAS core reconstructed within the array) and zenith angle ($<40^\circ$), 
chosen to assure best and constant reconstruction accuracies,
result in an exposure of $2 \cdot 10^{13}\,$m$^2 \cdot$s$ \cdot$sr. 
Figure~\ref{2dim} displays the correlation of the two observables $N_{ch}$ and $N_\mu$.
This distribution is the basis of the following analysis, since 
it contains all the experimental information required for reconstructing 
energy and mass of the cosmic rays: 
the higher the energy  of the primary cosmic ray the larger the total particle number. 
The fraction of muons of all charged particles at observation level is characteristic 
for the primary mass: 
showers induced by heavy primaries start earlier in the atmosphere and the higher 
nucleon number leads to a relatively larger muon content at observation level. 
KASCADE-Grande measures the particle number at an atmospheric depth well beyond the 
shower maximum, where the electromagnetic component already becomes reduced.
Thus, electron-rich EAS are generated preferentially by light primary nuclei and
electron-poor EAS by heavy nuclei, respectively.

However, a straightforward analysis is hampered by the shower-to-shower 
fluctuations, i.e. by the dispersion of the muon and electromagnetic particle numbers  
for a fixed primary mass and energy.
In addition, cosmic rays impinging on the atmosphere under different zenith angles 
show a varying, complicated behavior due to the non-uniform mass and density 
distribution of the air.
Therefore, the absolute energy and mass scale have to be inferred from comparisons of 
the measurements with Monte Carlo simulations. 
This creates additional uncertainties, since the physics of the relevant particle 
interactions is not completely tested by man-made accelerator experiments. 
The uncertainties imposed by the hadronic interaction models are more relevant 
for composition analyses than for energy measurements.
Hence, our strategy is to separate the measured EAS in electron-poor and electron-rich 
events as representatives of the heavy and light primary mass groups, 
similar to the analysis presented in Ref.~\cite{kas-muo}. 
The shape and structures of the resulting energy spectra of these individual mass groups
are much less affected by the differences of the various hadronic interaction 
models than the relative abundance.

 \begin{figure}[!t]
  \centering
  \includegraphics[width=2.4in]{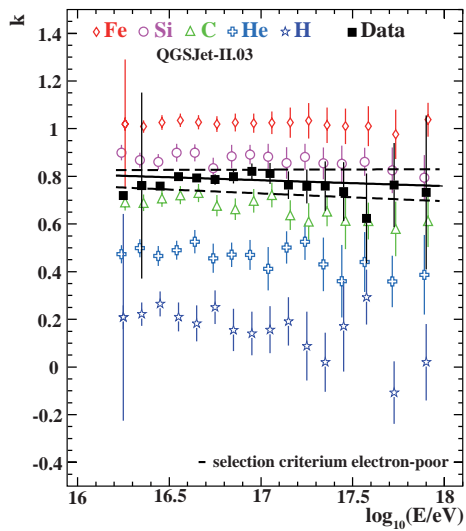}
  \caption{Evolution of the $k$ parameter as a function of the 
  reconstructed energy for experimental data compared with simulations of primary
	masses for the angular range 0-24$^\circ$. The error bars assign statistical as 
	well as reconstruction uncertainties of $k$.
	The line displays the chosen energy dependent $k$-values for separating the mass groups, 
	where the dashed lines assign the uncertainty of the selection.}
  \label{kparam}
 \end{figure}
 \begin{figure}[ht]
  \centering
  \includegraphics[width=3.0in]{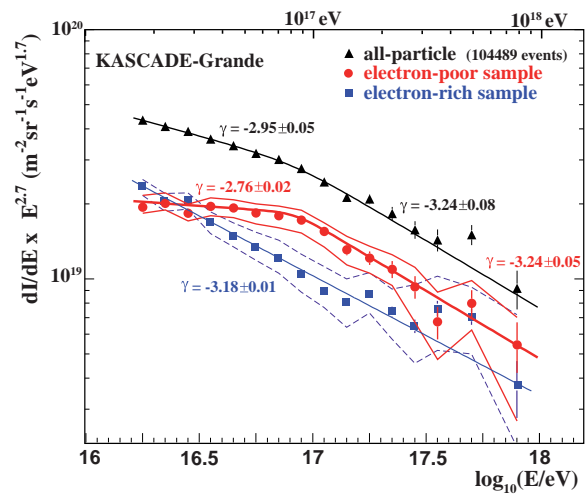}
  \caption{Reconstructed energy spectrum of the electron-poor and electron-rich 
  components together with the all-particle spectrum for the 
  angular range 0-40$^\circ$. The error bars show the statistical uncertainties; the bands 
  assign systematic uncertainties due to the selection of the subsamples. 
  Fits on the spectra and resulting slopes are also indicated.}
  \label{kspectra}
 \end{figure}
%
As a consequence of the considerations above, the energy and mass 
assignment of individual events is achieved by using 
both observables $N_{ch}$ and $N_\mu$, as well as their correlation.
The following equation is motivated by discussions of hadronic air showers in 
reference~\cite{matthews}, with the basic idea that the total number of secondary 
particles at observation level is related to the primary energy while the energy 
sharing between the electromagnetic and the hadronic (i.e.~muonic) shower components 
is related to the primary mass. 
Therefore, the primary energy $\log_{10}(E)$ is assumed to be proportional to the shower 
size $\log_{10}(N_{ch})$ with a correction factor that accounts for the mass dependence 
by making use of the measured ratios of shower sizes $\log_{10}(N_{ch}/N_{\mu})$:
{
 \begin{eqnarray}
  \log_{10}(E/GeV)&=&[a_H+(a_{Fe}-a_H) \cdot k]\cdot \log_{10}(N_{ch}) \nonumber\\
  &&{+}\: b_H+(b_{Fe}-b_H) \cdot k
  \label{equn1}
 \end{eqnarray}              %
 \begin{equation}
    k = \frac{\log_{10}(N_{ch}/N_{\mu})-\log_{10}(N_{ch}/N_{\mu})_H}
        {\log_{10}(N_{ch}/N_{\mu})_{Fe}-\log_{10}(N_{ch}/N_{\mu})_H}
  \label{equn2}
 \end{equation} 
}
\noindent with 
$\log_{10}(N_{ch}/N_{\mu})_{H,Fe} = c_{H,Fe} \cdot \log_{10} (N_{ch}) + d_{H,Fe}$.
The parameter $k$ takes into account both the average differences in the $N_{ch}/N_\mu$ 
ratio among different primaries with same $N_{ch}$ as well as the shower to shower 
fluctuations for events of the same primary mass. 
%
%
The exact form of the equation is optimized for the experimental situation of 
KASCADE-Grande and the free parameters~\cite{espec} are determined by 
Monte Carlo simulations 
\footnote{Simulations include the full air shower development in the atmosphere 
and the response of the detector~\cite{geant}. 
The EAS were generated using CORSIKA~\cite{cors} and the
models FLUKA~\cite{fluka} and QGSJet~II.03~\cite{qgs} in the energy range from 
$10^{15}\,$eV to $3 \dot 10^{18}\,$eV for five different representative mass 
groups: H, He, C, Si and Fe with about $353.000$ events per primary.}.
They are defined independently for 5 different zenith angle intervals 
of equal exposure (the upper limits of $\theta$ are $16.7^\circ$, $24.0^\circ$, $29.9^\circ$, 
$35.1^\circ$, and $40.0^\circ$) to take into account the shower attenuation. 
Data are combined only at the very last stage to reconstruct the final 
energy spectrum.
The $N_{ch}$-$N_\mu$-correlation of individual events is incorporated in calculating $k$,
which serves now as mass sensitive observable.  
Fig.~\ref{kparam} shows the evolution of $k$ as a function of the reconstructed energy
for the first two zenith angle bins, where a similar behavior is observed for all 
angular ranges. 
The error bars include statistical as well as reconstruction uncertainties of the 
$k$-parameter. 
The width of the $k$ distributions decreases slightly for increasing energy and 
amounts, at $100\,$PeV, to about $\pm0.2$, $\pm0.15$, $\pm0.4$ for H, Fe and 
data, respectively. 

The $k$-parameter is used to separate the events into different samples.
The line in Fig.~\ref{kparam} separates the electron-poor (heavy) group, 
and is defined by fitting the $k_{ep}(E)=(k_{Si}(E)+k_{C}(E))/2$ distribution.
The dashed lines represent the uncertainties in defining this energy dependent 
selection cut. 
The resulting spectra are shown in Fig.~\ref{kspectra}, where the band indicates 
changes of the spectra when the cut is varied within the dashed lines shown 
in Fig.~\ref{kparam}.
The energy resolution for an individual event is better than 25\% over the entire energy 
range and the all-particle spectrum is reconstructed within a total systematic 
uncertainty in flux of 10-15\%~\cite{espec} 
\footnote{The resulting spectra for the present analysis are not corrected for 
reconstruction uncertainties. But, more detailed investigations~\cite{espec} have shown 
that the effects are smaller than the estimated uncertainty on the flux of 10-15\%.
In addition, the absolute energy scale depends on the used hadronic interaction model, e.g.~for 
EPOS~v1.99~\cite{epos} a 10-15\% lower flux in the all-particle spectrum is obtained.}. 

The reconstructed spectrum of the electron-poor events shows a distinct knee-like 
feature at about $8 \cdot 10^{16}\,$eV. 
Applying a fit of two power laws to the spectrum interconnected by a smooth 
knee~\cite{bayes02} results in a statistical significance of $3.5\sigma$ that the 
entire spectrum cannot be fitted with a single power law.
The change of the spectral slope is $\Delta \gamma = -0.48$ from
$\gamma = -2.76 \pm 0.02$  to $\gamma = -3.24\pm0.05$ with the break position at
$\log_{10}(E/eV)=16.92\pm0.04$. 
Applying the same function to the all-particle spectrum results in a statistical
significance of only $2.1\sigma$ that a fit of two power laws is needed to 
describe the spectrum. 
Here the change of the spectral slope is from
$\gamma = -2.95 \pm 0.05$  to $\gamma = -3.24\pm0.08$, but with the break 
position again at $\log_{10}(E/eV)=16.92\pm0.10$. 
Hence, the selection of heavy primaries enhances the knee-like feature that is already 
present in the all-particle spectrum. 
The spectrum of the electron-rich events (light and medium mass primaries) 
is compatible with a single power law with slope index $\gamma = -3.18\pm0.01$. 
However, a recovery to a harder spectrum at energies well above $10^{17}\,$eV 
cannot be excluded by the present data. 
This finding is of particular interest and needs more detailed investigations with 
improved statistics in future.
 \begin{figure}[!t]
  \centering
  \includegraphics[width=3.0in]{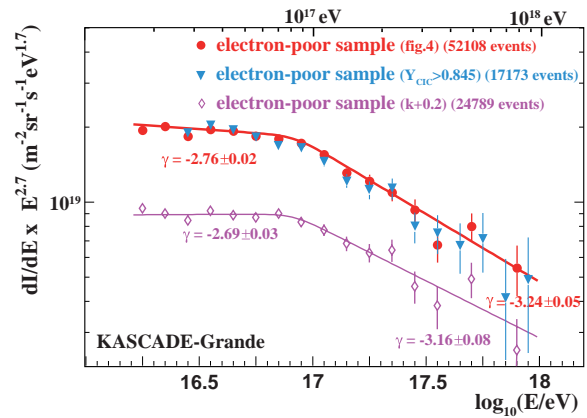}
	\caption{Energy spectra of electron-poor (heavy) event samples obtained by different 
	selection and reconstruction criteria. The original spectrum from Fig.~\ref{kspectra} is compared
	with the spectrum from a more selective cut in the $k$-parameter and
	with the spectrum obtained by using the $Y_{CIC}$-parameter for 	
	selecting the electron-poor events (see text).}
	\label{yspectra}
 \end{figure}
%

The main result, i.e.~the knee-like structure in the spectrum of electron-poor events, is 
validated in the following by various cross-checks (Fig.~\ref{yspectra}).
Variations of the slopes of the selection cut, as well as parallel shifts 
of the cut lines have shown that the spectral form, i.e.~the knee-like structure of 
the electron-poor event sample, is retained. 
By shifting $k$ to larger values the fraction of heavy primaries in the sample
is enriched. Interestingly, we found that the slope index of the spectrum is not significantly 
changing beyond the break, but gets systematically harder at lower energies . 
The position of the break remains constant, indicating that 
the heaviest primaries in the sample dominate the spectral form.
An example of a spectrum obtained by such a variation of the selection cut is 
shown in Fig.~\ref{yspectra}. 

A systematic uncertainty possibly affecting the interpretation of the data 
is related to the attenuation of the particle numbers in the atmosphere. 
So far, the attenuation given by the EAS simulations is taken into account.
For validation, an independent analysis is performed where the correction 
for attenuation, i.e.~for the zenith angular dependence, is based on the 
measured events, and not on simulations.
The correction parameters are obtained by applying the 
Constant Intensity Cut Method (CIC)~\cite{CIC} to the two observables independently.
This procedure allows the data collected from different zenith 
angles to be combined in a model independent way. 
The shower size ratio $Y_{CIC} = \log_{10}{N_\mu^\prime} / \log_{10}{N_{ch}^\prime}$ is 
calculated, where $N_\mu^\prime$ and $N_{ch}^\prime$ are the shower sizes 
corrected for attenuation effects in the atmosphere in such a way that they 
correspond to the shower sizes at a certain reference zenith angle. 
In order to check, in addition to the attenuation correction, also reconstruction 
and selection uncertainties, we applied for this analysis more stringent
cuts, which increase the energy threshold and decrease the statistics of the 
event sample compared to the standard analysis. 
Now, $Y_{CIC}$ is used to separate the events into 
electron-rich and electron-poor subsamples. 
In contrast to the $k$-parameter, the $Y_{CIC}$-parameter is almost 
energy independent, where the energy of the individual 
events is again determined using eqn.~\ref{equn1}.
For direct comparison with the results obtained before, $Y_{CIC} > 0.845$ is chosen 
for selecting the electron-poor event sample. 
The reconstructed spectrum (see Fig.~\ref{yspectra}) 
obviously confirms the earlier finding of the knee-like structure, which is due to a 
decrease in the flux of the heavy component. 

Another source of systematic uncertainty is related to the hadronic 
interaction model. 
In the frame of QGSJet-II, the measured distributions in $k$ and $Y_{CIC}$ 
are in agreement with a dominant electron-poor composition for the entire energy range. 
Whereas the $Y_{CIC}$ and $k$ values themselves behave differently for other 
hadronic interaction models, the measured and simulated $Y_{CIC}$- and $k$-dependences 
on energy, and hence the shapes and structures of the resulting spectra, are similar 
\footnote{First analyses based on 
simulations with the hadronic interaction model EPOS~v1.99~\cite{epos} have confirmed 
the findings of the knee-like feature in the spectrum of the heavy component 
for both analysis approaches, though the relative abundance of the 
subsamples change considerably.}. 
Details will be discussed in a forthcoming paper, but it is not expected that 
the basic result of the present analysis changes.

Summarizing, by dividing KASCADE-Grande measured air-shower events in electron-rich and 
electron-poor subsamples, there is first evidence that at about $8 \cdot 10^{16}\,$eV 
the spectrum of the
heavy component of primary cosmic rays shows a knee-like break.
The spectral steepening occurs at an energy where the charge dependent knee 
of primary iron is expected, when the knee at about $3$-$5 \cdot 10^{15}\,$eV  
is assumed to be caused by a decrease in the flux of primary protons. 

The authors would like to thank the members of the
engineering and technical staff of the KASCADE-Grande
collaboration, who contributed to the success of the experiment.
KASCADE-Grande is supported
by the BMBF of Germany, the MIUR and INAF of Italy,
the Polish Ministry of Science and Higher Education (grant for the years 2009-2011),
and the Romanian Authority for Scientific Research UEFISCDI, 
(grants PNII-IDEI code 1442/2008 and PN 09 37 01 05).
JC.A., A.H. and M.F. acknowledge partial support from the DAAD-Proalmex program (2009-10),
JC.A. from CONACYT and the Consejo de la Investigaci\'on Cient\'ifica of 
the Universidad Michoacana.



\begin{thebibliography}{99}
\bibitem{kas-muo} 
T. Antoni et al. (KASCADE Collaboration), Astrop. Phys. {\bf 16} (2002) 373.
\bibitem{kas-unf} 
T. Antoni et al. (KASCADE Collaboration), Astrop. Phys. {\bf 24} (2005) 1.
\bibitem{peters}
B. Peters, Nuovo Cimento {\bf 22} (1961) 800.
\bibitem{hoera}
J.R.~Hoerandel, Astrop. Phys. {\bf 21} (2004) 241. 
\bibitem{kg-NIM10}
W.-D. Apel et al. (KASCADE-Grande Collaboration), NIM A {\bf 620} (2010) 202. 
\bibitem{kascade} 
T.~Antoni et al. (KASCADE Collaboration), NIM A {\bf 513} (2003) 429.
\bibitem{matthews} 
J. Matthews, Astrop. Phys. {\bf 22} (2005) 387.
\bibitem{espec}
M. Bertaina et al. (KASCADE-Grande Collaboration), Astrophys. Space Sci. Trans. {\bf 7} (2011) 229. 
\bibitem{geant}
CERN, GEANT 3.21, Detector Description and Simulation Tool CERN Program Library Long Writeup W5015 (1993).
\bibitem{cors}
D. Heck et al., Report FZKA 6019, Forschungszentrum Karlsruhe (1998).
\bibitem{fluka}   
A.~Fass\`o et al., Report CERN-2005-10, INFN/TC-05/11, SLAC-R-773 (2005).
\bibitem{qgs} 
S.S.~Ostapchenko, Nucl. Phys. B (Proc. Suppl.) {\bf 151} (2006) 143\&147;
S.~Ostapchenko, Phys. Rev. D 74 (2006) 014026. 
\bibitem{bayes02} 
T. Antoni et al. (KASCADE Collaboration), Astrop. Phys. {\bf 16} (2002) 245.
\bibitem{CIC}
J. Hersil et al., Phys. Rev. Lett. {\bf 6} (1961) 22.; 
D.M. Edge et al., J. Phys. A: Math. Nucl. Gen. {\bf 6} (1973) 1612.
\bibitem{epos}
K.~Werner, F.M.~Liu, T.~Pierog, Phys. Rev. C {\bf 74} (2006) 044902.

\end{thebibliography}
\end{document}